# Electron string ion sources for carbon ion cancer therapy accelerators

A.Yu. Boytsov[1], D.E. Donets[1], E.D. Donets[1], E.E. Donets[1], K. Katagiri[2], K. Noda[2],

D.O. Ponkin[1], A.Yu. Ramzdorf[1], V.V. Salnikov[1] and V.B. Shutov[1]

[1]*Joint Institute for Nuclear Research, Dubna, 141980, Russia*

[2]*National Institute of Radiological Science, Chiba, Japan*

The Electron String type of Ion Sources (ESIS) was developed, constructed and tested first in the Joint Institute for Nuclear Research.[1] These ion sources can be the appropriate sources for production of pulsed $C^{4+}$ and $C^{6+}$ ion beams which can be used for cancer therapy accelerators. In fact the test ESIS Krion-6T already now at the solenoid magnetic field only 4.6 T provides more than $10^{10}$ $C^{4+}$ ions per pulse and about $5 \cdot 10^{9}$ $C^{6+}$ ions per pulse. Such ion sources could be suitable for application at synchrotrons. It was also found, that Krion-6T can provide more than $10^{11}$ $C^{6+}$ ions per second at 100 Hz repetition rate, and the repetition rate can be increased at the same or larger ion output per second. This makes ESIS applicable at cyclotrons as well. As for production of $^{11}C$ radioactive ion beams ESIS can be the most economic kind of ion source. To proof that the special cryogenic cell for pulse injection of gaseous species into electron string was successfully tested using the ESIS Krion-2M.[2]

## I. INTRODUCTION

Cancer therapy with use of accelerated carbon ions is a new technologically advanced method of delivering radiation dose to the tumour while protecting surrounding healthy tissues. Moreover, carbon ion therapy can reach otherwise difficult to access deep-seated tumours and can be efficiently used for radioresistant tumours as in hypoxia. During last decades clinical use of carbon ion therapy was mostly developed in National Institute for Radiological Sciences (Chiba, Japan) and nowadays 6 clinics around the world use carbon ion therapy[3]. Significant potential exists to develop systems to enhance the effectiveness of carbon beam radiotherapy and we concentrate here on a related ion source developments carried out in Joint Institute for Nuclear Research (JINR, Dubna, Russia).

Electron String Ion Source (ESIS)[1] is the basically modified Electron Beam Ion Source (EBIS)[4] and they both were first proposed, developed, constructed, tested and used in JINR. In ESIS multiply reflected electrons, which in special conditions can form electron string (pure one component hot electron plasma) are used instead of direct electron beam which is used in EBIS. The effective number of electron reflections in electron string state is rather high. That allows using relatively low injected electron current to keep the string in a steady state and produce ion beams. The ion source Krion-2M was the first one where the electron string phenomenon was first discovered and used for production of multiply and highly charged ions.[1]

It was shown that number of the reflections and ion output of ESIS grow rapidly with increase of the source solenoid magnetic field. The new ESIS Krion-6T was recently developed as a prototype of the ion source for the NICA accelerator project[5] and appeared to be convenient also for production of $C^{4+}$ and $C^{6+}$ ions. According to the project the superconducting focusing solenoid of Krion-6T has to be used in the persistent mode of operation at 6 T magnetic field. And as a system of its active protection is not ready yet the ion source is used mostly with the solenoid magnetic field B = 4.6 T. The experiments on Krion-6T presented below were fulfilled mostly at this value of the focusing magnetic field.

In next Section we briefly describe experimental setup in its present status, while in the Section III main recent results on $C^{4+}$ and $C^{6+}$ ion beams production with Krion-6T ESIS are presented. In Section IV we discuss results of our recent studies towards production of a radioactive (positron emitting) carbon ion beams $^{11}C^{4+}$ and $^{11}C^{6+}$, which are planned for further carbon ion therapy simultaneously with on-line positron emission tomography (PET). In Section V we present results on multiply step ion extraction of produced carbon beams from ESIS. Finally we summarize results in Conclusions Section.

## II. EXPERIMENTAL SETUP

The electron-ion optics system of Krion-6T consists of the electron gun, consisting from electron emitter, dummy cathode and anode, situated in a fringe field of 1.2 meter superconducting solenoid at 1/20 of Bmax and the electron reflector, consisting from electron repeller and anode, situated on the other side of the solenoid in the same value of magnetic field. The electrons, emitted by the emitter, after many oscillations between the repeller and dummy cathode are collected on the anodes. The electron drift structure consisting of 26 drift tube sections is placed between the gun and the reflector along the axis of the solenoid magnetic field. The sections are 5 cm length and 4 mm inner diameter. The first five and the last two sections are placed on 78 K temperature terminal while the others 19 sections -on 4.2 K terminal. The main ion trap is placed inside the sections from 8$^{th}$ to 23$^{d}$ and on the sections 7 and 24 electrical potential barriers were applied to limit the ion trap in the axial direction. The working gas CH4 for production of carbon ions was transported under very low pressure into the fifth section. So, that when the electron string was switched on the ions were produced. To inject ions into the main ion trap the potential barrier on the 7$^{th}$ section was switched off while it was switched on on the second section. When the main trap was filled by ions the potential barrier was switched on again on the 7$^{th}$ section while it was "off" on the 2$^{nd}$ one. The ions were confined in the trap as long period of time as it was necessary for reaching desired charge state due to electron impact ionization. Then the produced ions were extracted from the trap in axial direction through the special orifice in the electron repeller and were directed to the Faraday cup or into the 2 m time of flight charge state analyzer. The electron-ion optics of the ESIS Krion-2M with 3 T solenoid is similar to that of Krion-6T but the gas injection section additionally equipped with a special cryogenic cell for pulse gas injection which can be useful for economic production of $^{11}$C radioactive ion beams. The short description of the cell will be done in Section IV.

## III. PRODUCTION OF $C^{4+, \, 6+}$ IONS BEAMS

Effective electron beam current density of electron string in Krion-6T usually reached several hundreds A/cm$^2$ at electron energy 5 – 7 KeV and at only about 10 mA electron injection current feeding the string while the space charge of string electrons per 1 m length of the ion trap reached 10 – 15 nC.

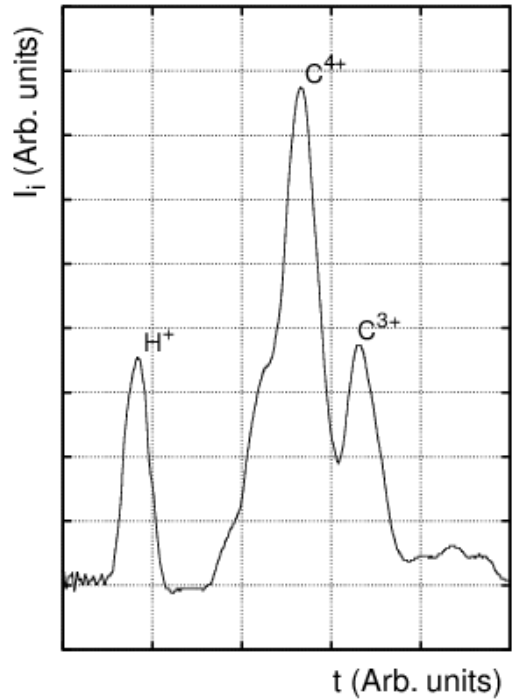

Fig. 1. The charge state spectrum obtained after 3 ms injection.

Such high the current density provided ionization of CH4 molecules carbon to $C^{4+}$ state in a maximum intensity in charge state spectra in 2-3 ms which was comparable with time of ion injection into the ion trap. At such conditions of $C^{4+}$ ion beam production we extracted ions just immediately after the end of the ion injection. In this case the charge state distribution can be regulated by changing CH4 pressure in the injection section. In Fig.1 the charge state spectrum obtained after 3 ms injection is presented. One can see that the positive ion charge of $C^{4+}$ ions was about half of the total ion charge. And the total ion charge was 14 nC per pulse in this case. That means that already now using the ESIS Krion-6T one can produce about $1,1 \cdot 10^{10}$ $C^{4+}$ ions per pulse, which is considered as sufficient for synchrotrons and it is expected that at higher magnetic field the ion output will be even higher.

Production of $C^{6+}$ ion beam requires essentially larger time of ion confinement. Taking into account ion losses from the ion trap due to electron heating during the confinement, one cannot expect so large $C^{6+}$ ion output as the $C^{4+}$ one. In Fig. 2 the charge state spectrum is presented for 3 ms ion injection into the ion trap and 18 ms ion confinement in the trap. One can see that the positive ion charge of $C^{6+}$ ions was about 57 % of the total 6 nC ion charge, i. e. about $4 \cdot 10^9$ $C^{6+}$ ions per pulse and this number reached about $5 \cdot 10^9$ ions/pulse at 30 ms ion confinement. This is about twice less than desired number in $1 \cdot 10^{10}$, but at higher solenoid magnetic field it can reach the number. One has to mention that the pulse power of the electron beam collected on the both gun and reflector anodes usually reached only 50 W and the average power in use an ESIS ion source at synchrotrons would be measured in mW.

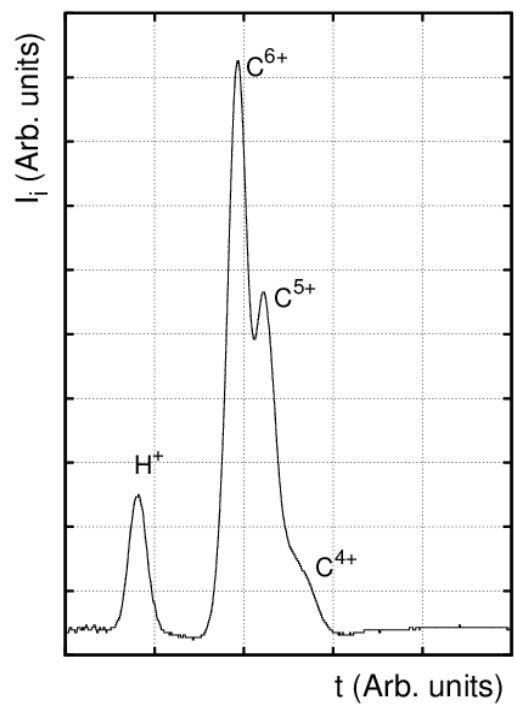

Fig. 2. The charge state spectrum for 3 ms ion injection and 18 ms ion confinement.

There is the project[6] to use a fast cycling cyclotron in the carbon ion beam cancer therapy. For the project $10^{11}$ $C^{6+}$ ions per second is required in 2 μs ion pulses with a repetition rate equal to approximately 400 Hz. These strong requirements are still out of existing ion sources capabilities, however, serious efforts have been initiated recently towards reaching the desired parameters with use of ESIS[7] and EBIS[8,9] ion sources.

We continue this studies with use of our new Krion-6T ESIS which provides adequate capabilities. In Fig. 3 the carbon ion charge state spectrum, obtained using the Krion-6T, is presented for 1.5 ms ion injection and 8 ms ion confinement. Here the positive ion charge of $C^{6+}$ ions is about 28 % of the total ion charge, which in this case was 6.3 nC per pulse. This result shows that with this ion source one can produce about $1.8 \cdot 10^{11}$ ions per second at 100 Hz repetition rate and the rate can be increased with the same or even larger ion output.

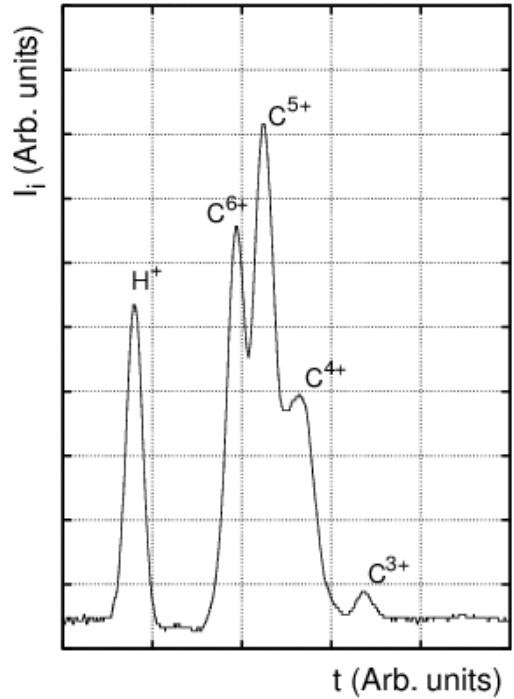

Fig. 3. The charge state spectrum for 1.5 ms ion injection and 8 ms ion confinement.

## IV. TOWARDS PRODUCTION OF $^{11}C^{4+, 6+}$ IONS BEAMS

Ion beams of radioactive $^{11}C$ carbon of the same or comparable intensity with that of the stable carbon ion beams could essentially improve patient treatments due to possibility of simultaneous area irradiation and dose controls[10]. As $^{11}C$ half life is only about 20 min. an effective conversion of $^{11}C$ produced in a nuclear reaction to $^{11}C^{4+}$ or $^{11}C^{6+}$ ion beams in ion source is of a great importance. And as the time of carbon confinement in the ESIS ion trap for production for example $C^{4+}$ ions is less than 5 ms it is necessary to use a corresponding short pulse injection of the radioactive CH4 into the ion source electron string for the effective conversion.

To provide the pulse injection the special cryogenics methane keeping cell has been developed[2] and tested with the ESIS Krion-2M. The cell is schematically drawn in Fig. 4. The cell jacket (1) is placed on the source 78 K terminal just near the injection section (2) of the drift tube and it is connected with the section via a small short pipe. The metallic rod (4) connected with the source 4.2 K terminal is placed on the axis of the cell. The rod is covered by a thin electric and thermal insulation layer which is covered by an electric conductive layer.

In the test experiments the working gas (stable CH4) was transported to the cell by means of another pipe from an external reservoir. Inside the cell the gas immediately was frozen on the rod surface. When the surface was pulse heated up to approximately 40 K by pulse electric current through the conducting layer, all the frozen methane evaporated and its small part penetrated into the injection section. After the pulse the surface was cooled down again frizzing again all the rest methane for next injection pulses. In the injection section the injected methane was ionized in the electron string (3) and ions were accumulated in the source main ion trap for production of a pulse ion beam, which can be measured.

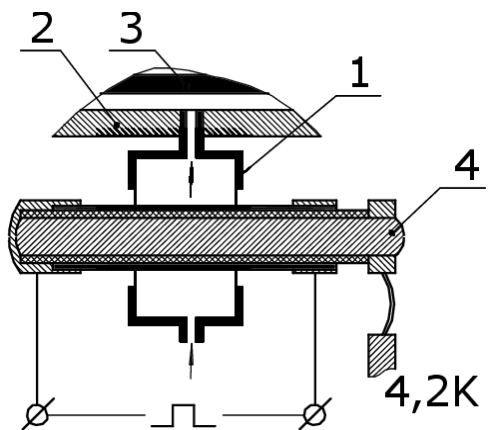

Fig. 4. The schematic drawing of the cryogenic sell.

We studied pulse properties of the methane keeping sell shifting the start of ion accumulation from the start of 2 ms heating pulse and measuring the corresponding accumulated positive ion charge and calculating its derivative. The result is presented in Fig. 5.

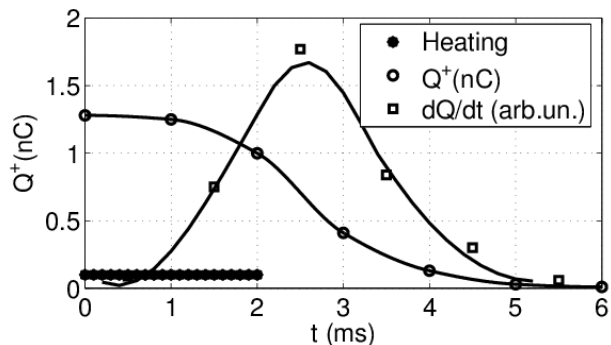

Fig. 5. Total accumulated ion charge and its derivative versus the delay time.

One can see that the pulse property of the cell quite good corresponds to the period of time necessary for $C^{4+}$ ions production from CH4 in the Krion-6T ion source and moreover for $C^{6+}$ ions.

To measure the conversion efficiency of methane to $C^{4+}$ ions we put a definite number of methane molecules into the external reservoir, froze them on the cell rod and then in series of 5 ms pulses injected into the Krion-2M source electron string, produced $C^{4+}$ ions and measured the total number of the ions per initial number of methane molecules. We found out that the conversion efficiency reached several percents, which can be acceptable for use of $^{11}C$ ion beams in cancer therapy.

**V. MULTIPLY STEP ION EXTRACTION**

As various accelerators are sensitive to different times of ion beam injection the multiply step ion extraction has been developed for ESIS ion sources. At this extraction the potential "bottom" of the source main ion trap is lifted up step by step and every step can be made different in height and in time. In different cases space distributions of ions in the string electrons space charge potential well can be different and the developed mode of ion extraction can provide in any case the necessary ion pulse shape. For example using one step extraction one can get 8 us ion pulse suitable for one turn injection synchrotrons. And in Fig. 6 (1) one can see the shape of carbon ion current pulse obtained with multiply step ion extraction which could be used for multiply turn injection into the NIRS synchrotron. Fig. 6 (2) represents even longer, about 300 us, ion current pulse suitable, for example, for time of flight charge state analysis at low ion current.

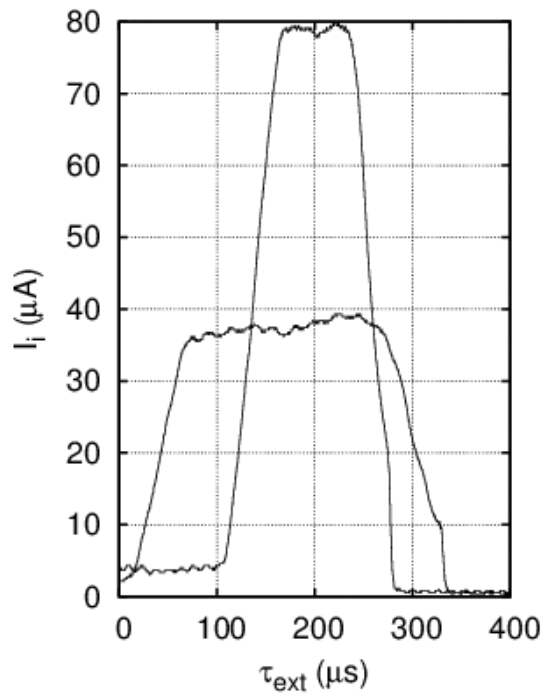

Fig. 6. The examples of extracted ion pulses.

In Fig. 7 about 2 us ion current pulse is presented obtained at ion extraction from a part of the main ion trap which shows a possibility using of the ESIS type ion source on the cyclotron accelerator complex.[6]

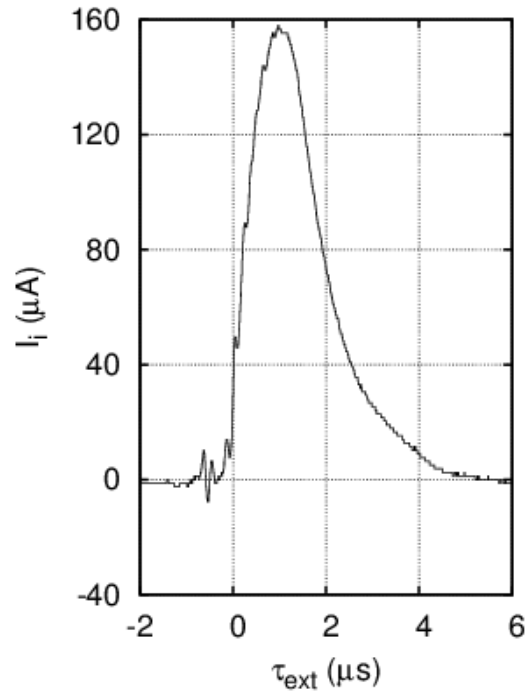

Fig. 7. The extracted ion pulse from a part of the main ion trap.

## VI. CONCLUSIONS

It was experimentally shown that an ESIS ion source can provide more than $10^{10}$ $C^{4+}$ ions per pulse, which is more than necessary for cancer therapy synchrotrons. It was also shown that one can expect the same number of $C^{6+}$ ions per pulse and therefore possibility of a considerable simplification of synchrotron injectors. An ESIS ion source equipped with the cryogenic methane keeping sell could be also the appropriate source for using radioactive $^{11}C$ ions.

## ACKNOWLEDGMENTS

The authors are thankful to Dr. G.V. Trubnikov and to Dr. A.V. Butenko for the kind support and interest during developments of the ESIS Krion-6T.